\documentstyle[11pt,newpasp,twoside,epsf]{article}
\def\edcomment#1{\iffalse\marginpar{\raggedright\sl#1\/}\else\relax\fi}
\reversemarginpar

\def\lta{{\>\rlap{\raise2pt\hbox{$<$}}\lower3pt\hbox{$\sim$}\>}}
\def\gta{{\>\rlap{\raise2pt\hbox{$>$}}\lower3pt\hbox{$\sim$}\>}}

\begin{document}
\title{The Metallicity of $0.5<z<1$ Field Galaxies}
 \author{C. Marcella Carollo}
\affil{Columbia University,  Department of Astronomy, New York, NY 10027, USA}
\author{Simon J. Lilly}
\affil{Herzberg Institute of Astrophysics, National Research Council, Victoria V9E 2E7, Canada}
\author{Alan Stockton}
\affil{Institute for Astronomy, University of
              Hawaii, Honolulu, HI 96822, USA}

\begin{abstract}
We are undertaking a program to measure emission line ratios in a
selected sample of $0.5 < z < 1.0$ CFRS starforming galaxies in order
to compute their Interstellar Medium metallicities. The latter are
derived by means of the empirically-calibrated $R_{23}$ estimator
introduced by Pagel et al.\ (1979).  Here we focus on a subsample of
15 galaxies with $L_{H\beta} > 1.2 \times 10^{41}$ erg s$^{-1}$.  In
addition to the optical CFHT spectra discussed in Carollo \& Lilly
(2001), where a preliminary account of this work can be found, in this
paper we also refer to new $J$-band Keck spectroscopy of the $H\alpha$
and [NII]6583 lines for several objects. These lines allow us
to break the degeneracy between low ($\lta 0.1Z_\odot$) and high
($\gta 0.5Z_\odot$) metallicity of the $R_{23}$ estimator, and put on
solid ground the finding that the $0.5 < z < 1.0$ blue galaxies in the
high-$L_{H\beta}$ sample are significantly metal-enriched ($Z \gta
0.3Z_\odot$ up to $Z_\odot$).  Therefore, our results do not support previous
suggestions in which the $0.5 < z < 1.0$ starforming galaxies are dwarf
galaxies brightened by large bursts of star-formation, but support
instead a picture where at least a significant fraction of them
evolves to become today's massive metal-rich systems.
\end{abstract}

\section{Introduction}

The Canada-France Redshift Survey
(CFRS; Lilly et al.\ 1995a) and other
similar surveys have shown  evolutionary changes in the field galaxy
population over the redshift interval $0 < z < 1$.
At $z \sim 1$ a substantial population  of  $L
\sim L^*$, blue  galaxies  
with small sizes and irregular morphologies (Brinchmann et al.\ 1998,
hereafter B98; Lilly et al.\ 1998; hereafter L98) appears, a population
that is largely absent in the local Universe. 

Studies that have provided e.g., the morphologies and sizes of the
$z\sim1$ field blue galaxies (e.g. B98; L98), and their internal
kinematics have suggested that these blue galaxies have generally the
irregular morphologies (B98), small sizes (3-5 $h_{50}^{-1}$ Mpc, L98)
and low velocity dispersions $\sigma < 100$ kms$^{-1}$ (Guzman et al.\
1997; Vogt et al.\ 1997; Mallen-Ornelas et al.\ 1999) that are usually
associated in the local Universe with galaxies 2-3 magnitudes further
down the luminosity function. These results support at first sight a
scenario where the $z\sim1$ blue galaxies are dwarf galaxies
substantially brightened by a burst of star formation.  There are
still, however, major gaps in our knowledge of galaxies in the $0.5 <
z < 1.0$ redshift regime, and it is possible that this interpretation of
the morphologies and the kinematic data is incorrect.  Not least,
the possibility that the small-blue-irregular galaxies are the cores
of more massive galaxies cannot be ruled out at this stage. For
example, the $K$-band luminosities of typical blue galaxies at $z \sim
0.8$ are comparable to those of today's massive galaxies, leading to
suggestions that at least some of these blue objects may be indeed the
progenitors of massive systems (Cowie et al.\ 1996).

Detailed astrophysical diagnostics are needed to discriminate between
the two scenarios.  The metallicity of the interstellar medium (ISM)
of distant starforming field galaxies is particularly important, both
as a general indicator of the evolutionary state of these systems, and
as a constraint on their possible present-day descendants. No
information has been available about the ISM metallicities of galaxies
in the $0.5 < z < 1.0$ redshift regime.  We have therefore begun a
program of systematic emission line spectroscopy of CFRS galaxies to
determine their ISM metallicity, and to study how the metal content in
these systems correlates with galaxy luminosity, star-formation rate,
structure and morphology.  Our program uses the $R_{23} = ([OII]3727 +
[OIII]4959+5007)/H\beta$ metallicity estimator (Pagel et al.\ 1979;
see also Kobulnicky et al.\ 1999 for a detailed description of the
potential accuracy that can be obtained in using $R_{23}$ to measure
metallicities of unresolved galaxies at high redshifts).  
The lines required to measure $R_{23}$ are shifted into the 5000 $\AA$
- 1 $\mu$m wavelength region for $0.5 < z < 1$.  A reversal in
$R_{23}$ occurs however at $Z \sim 0.3 Z_{\odot}$ due to cooling
effects (so a low- and a high-metallicity solution are associated with
most values of $R_{23}$), and the [OIII]5007/[NII]6584 ratio is
required to break this degeneracy (Kobulnicky et al.\ 1999).  Our
program therefore includes supplementary spectroscopy of H$\alpha$, [NII]6584
and [SII]6717,6731 lines (which, for $0.5 < z < 1$, are shifted into
the near-infrared $J$-band) for reddening determinations, AGN
discrimination and breaking of the $R_{23}$-degeneracy with
metallicity through the [OIII]/[NII] ratio.

In this paper we use deep $0.5 < \lambda < 1.0 \mu$m multi-object
spectrophotometry collected at the Canada-France-Hawaii Telescope in
March 2000 (already published in Carollo \& Lilly 2001), and refer to
complementary $J$-band spectroscopy that we have in the meanwhile
acquired with NIRSPEC at Keck for five objects in the optical sample (in
preparation). We use $H_o=50$ Mpc/Km/s and $q_o=0.5$

\begin{figure}
\plottwo{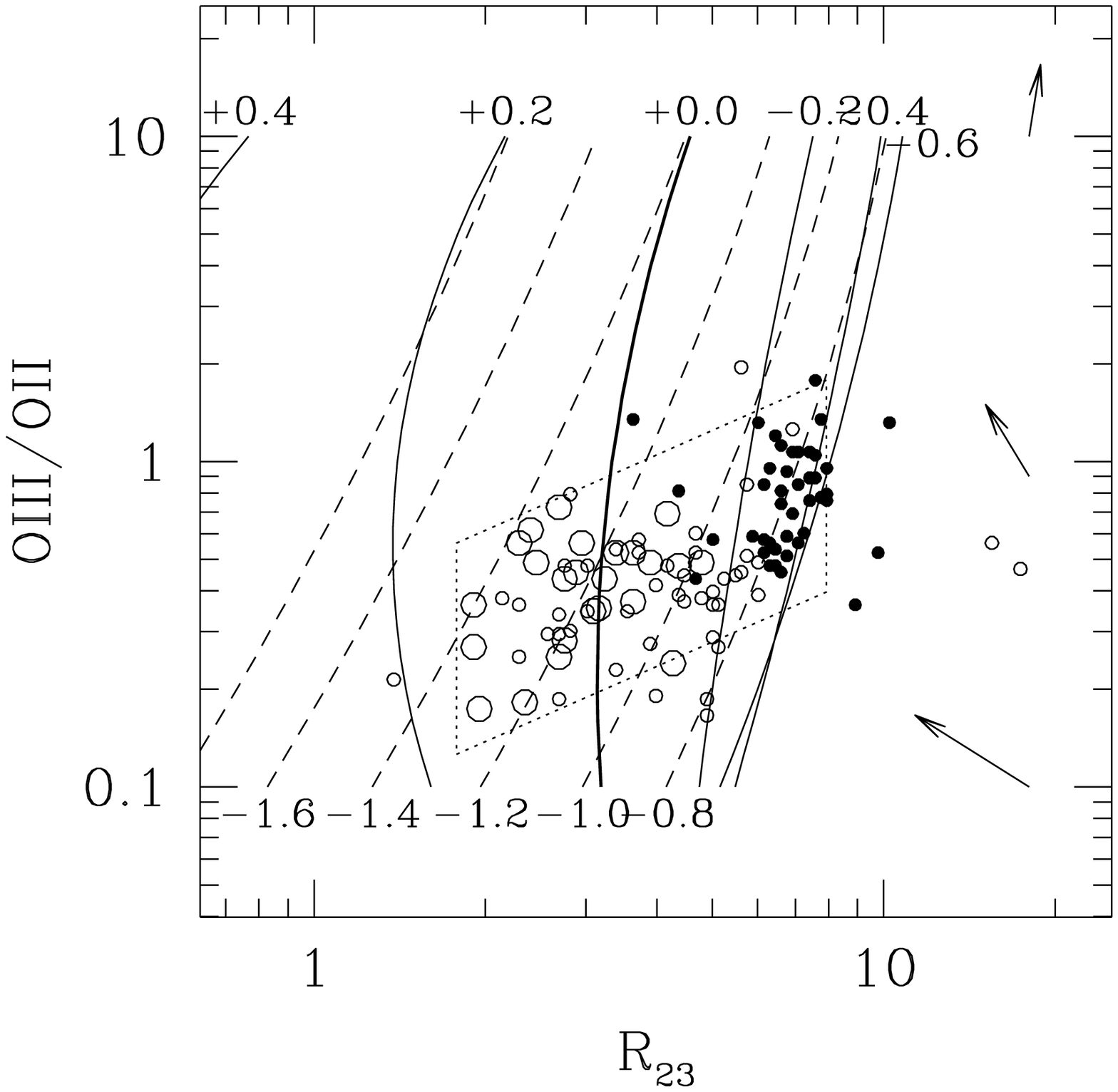}{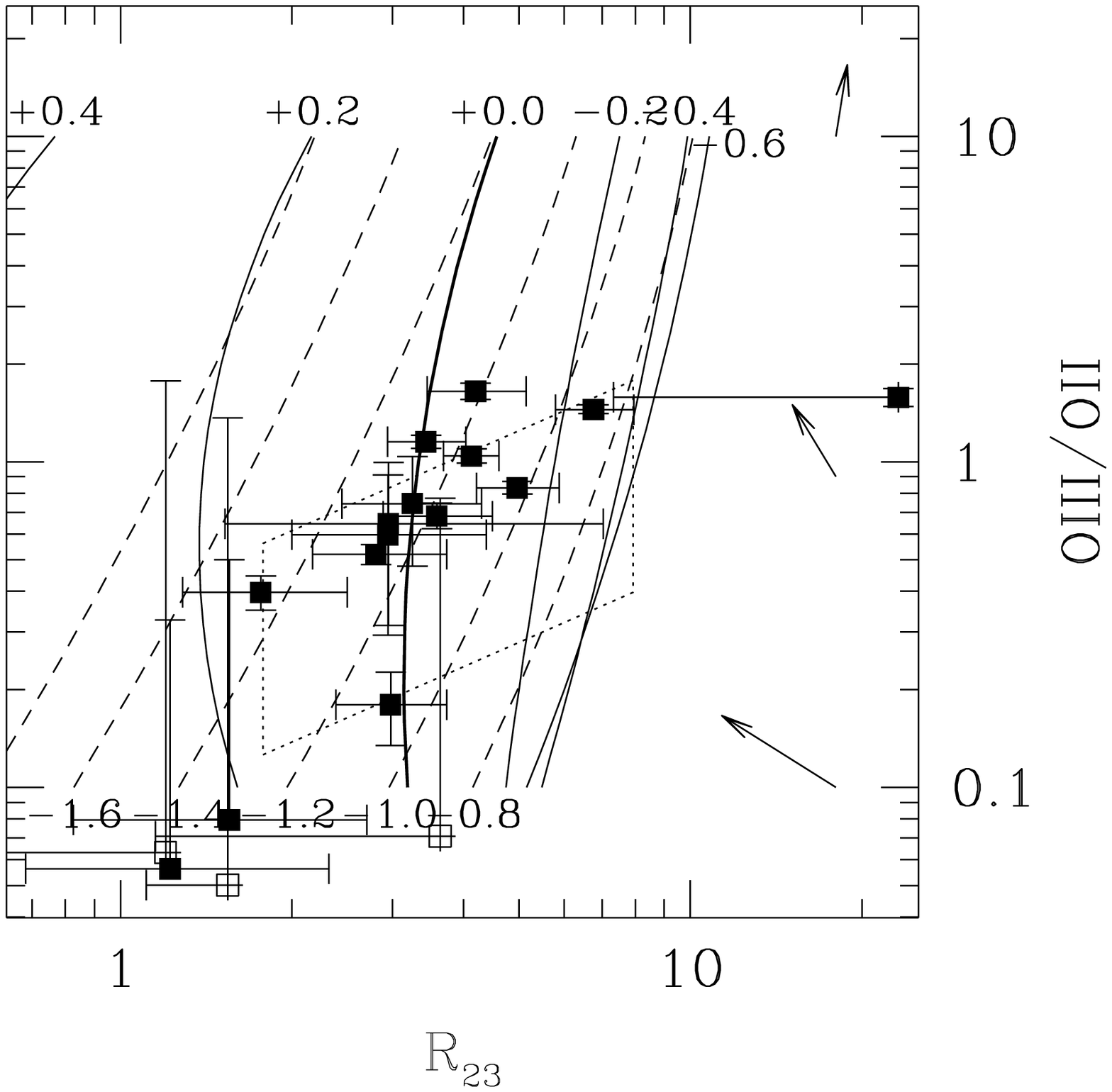}
\caption{The [OIII]5007/[OII]3727 versus $R_{23}$ relation for the
local field sample of Jansen et al. (left, panel a) and the
L$_{H\beta} > 1.2\times 10^{41}$ ergs s$^{-1}$ CFRS galaxies of our
sample (right, panel b). Symbols are explained in the text. }
\end{figure}

\section{The Sample}

The target galaxies were selected from the CFRS to have, in addition
to the original $I_{AB} < 22.5$ photometric selection, an [OII]3727
flux $ > 7 \times 10^{-17}$ erg s$^{-1}$cm$^{-2}$, so as to ensure
strong enough lines for a measurement of the $R_{23}$ parameter. 34
such objects were observed in our first CFHT observing run.  The H$\beta$
luminosity is however likely to be better related to the
star-formation rate than the [OII] flux.  Unfortunately, H$\beta$
measurements were not available beforehand, so an approximation to
an $H\beta$-luminosity selected sample could only be constructed a
posteriori.  The current optical analysis is therefore based on the 15 out of 34 objects
with H$\beta$ luminosities above $L_{H\beta} > 1.2
\times 10^{41}$ erg s$^{-1}$.  This high-$L_{H\beta}$ selection well
samples the `blue' starforming CFRS galaxy population.  Three high
redshift galaxies in the original sample of 34 had upper limits to
their H$\beta$ luminosities which were above our $L_{H\beta} = 1.2
\times 10^{41}$ erg s$^{-1}$ threshold, possibly putting these systems
into the high-$L_{H\beta}$ sub-sample that is discussed in this paper.
The line ratios for these galaxies are necessarily uncertain; however,
we have included these systems in the figures for completeness, and
assigned to them the range of line ratios that they would have if
their H$\beta$ luminosities were indeed above our threshold. We have
however identified these three objects with different symbols, as a
reminder that they may well not belong to the high-$L_{H\beta}$
sub-sample.

\section{Results and Discussion}

Figures 1a and 1b show the [OIII]5007/[OII]3727 versus $R_{23}$
relation for the local field sample of Jansen et al.\ (2000) and the
high-$L_{H\beta}$ CFRS galaxies of our sample, respectively.  In the
figures, the solid and dashed lines are lines of constant metallicity.
From left to right, the dashed lines indicate increasing metallicities
from 0.02$Z_\odot$ to 0.2$Z_\odot$. At $Z \sim 0.3Z_\odot$, the
reversal of $R_{23}$ occurs, and the curves of constant metallicity
then follow the solid-line sequence in which $Z$ rises up to about
3$Z_\odot$ from right back to left. The $Z = Z_\odot$ curve is
highlighted with a thicker linewidth.  In order to appropriately
compare the local and the high--$z$ samples, the local Jansen et al.\
galaxies which have L$_{H\beta} > 1.2\times10^{41}$ ergs s$^{-1}$ are
identified with large circles; local galaxies with smaller $H\beta$
luminosities are represented by small circles. Furthermore, in Figure
1a, filled symbols represent objects which have a flux ratio
$[OIII]5007/[NII]6584 > 2$ and thus metallicities $Z \lta 0.5
Z_\odot$, and empty symbols indicate objects with higher
metallicities, as indicated by a flux ratio $[OIII]5007/[NII]6584 < 2$
(see Edmunds \& Pagel 1984). In Figure 1b, the high--$z$ galaxies are
represented by the filled squares, with the exception of the three
objects which may or may not be in the sample on account of their
H$\beta$ upper limits (empty squares). The fiducial dotted-line box in
both panels encompasses the bulk of the Jansen's galaxies.  The arrows
represent the direction in which the points of the diagram would shift
due to reddening by dust (as described by Cardelli et al.\ 1989); the
length of the arrows refer to an $E(B-V)=0.3$ magnitudes at $z=0.7$.

The high redshift galaxies fall in the $R_{23}$ versus [OIII]/[OII]
plane in locations that are occupied by galaxies in the local Jansen
et al.\ (2000) sample.  Furthermore, most of the high redshift sample
selected to have $L_{H\beta} > 1.2 \times 10^{41}$ erg s$^{-1}$
occupies the same restricted location in the $R_{23}$ versus
[OIII]/[OII] plane as do the objects in the local Universe with
similarly high $H\beta$ luminosities.  At both epochs, galaxies
selected to have the same H$\beta$ luminosities exhibit the same range
of $R_{23}$ and [OIII]/[OII].  Two objects have, within the error
bars, $R_{23} \sim 7$, the value that is non-degeneratively associated
with intermediate metallicities $Z \sim 0.3 Z_{\odot}$. These two
objects were excluded from the discussion in Carollo \& Lilly (2001)
due to their [NeIII]3869 emission that made them good candidates for
AGN activity.  The $Z$-degenerate $R_{23}$ values that are measured
for the remaining high-$z$ galaxies indicate either rather low ($\lta
0.1Z_\odot$) or rather high ($\sim Z_\odot$) metallicities for these
systems.

\begin{figure}
\plotone{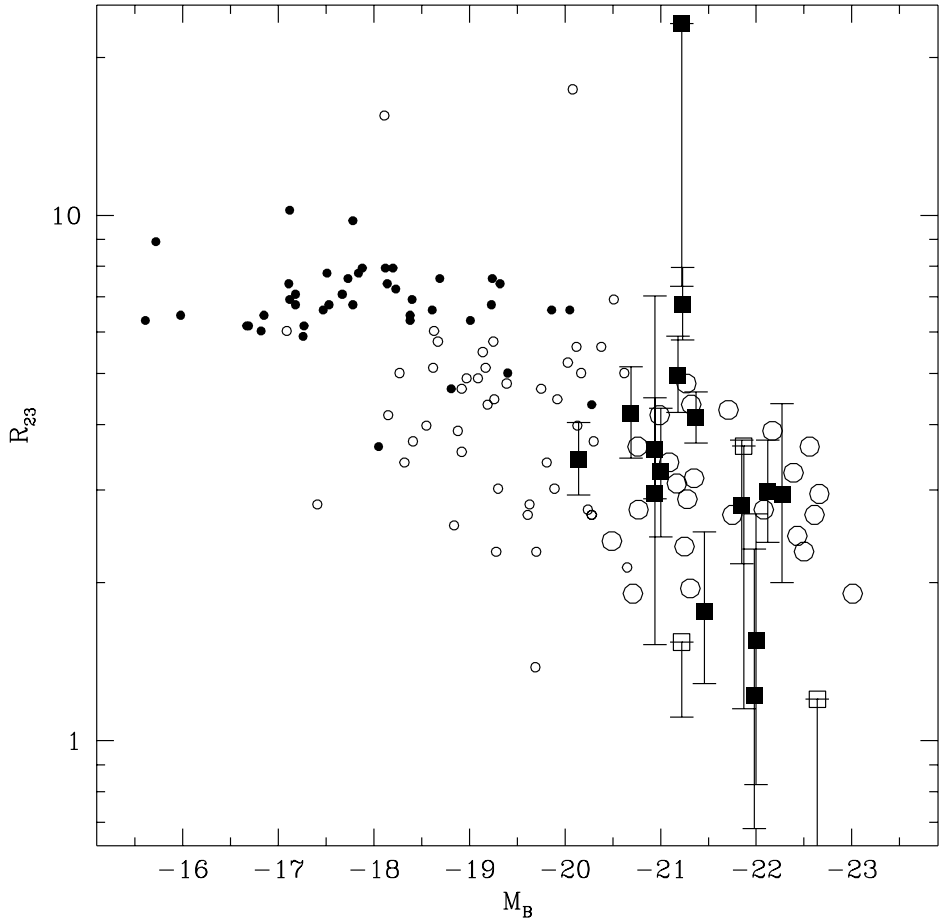}
\caption{The $R_{23}$ versus absolute $B$ magnitude  relation for
the galaxies in the local universe (circles) and those in the $0.5 < z
< 1$ redshift regime (squares).  Symbols are as in
Figures 1a and 1b, and are explained in the text.}
\end{figure}

The similarity between most of the high-redshift galaxies and the
low-redshift sample is further illustrated in Figure 2, which shows
the relationship between $R_{23}$ and continuum luminosity
M$_B$. There is a startling similarity between the high-$z$ and the
local galaxies on this diagram.  There is little evidence for an
evolutionary change in the relationship between metallicity (as
estimated from $R_{23}$) and the line and continuum luminosities for
high-$L_{H\beta}$ field galaxies between $z \sim 0$ and $z \sim 1$.
The two [NeIII] emitters are the only two systems out of 15 which may
indicate a different relationship, i.e. one produced by substantial
brightening of lower $Z$ systems.

$J$-band spectroscopy is available for five of the 15 galaxies,
including the two [NeIII] emitters. These new data provide the
intensity of the [NII]6584 emission line, and enable us to compute the
degeneracy-breaking [OIII]5007/[NII]6584 ratio.  Three of the five
objects have [OIII]5007/[NII]6584 ratios that undoubtedly place them
on the high-metallicity branch of the $R_{23}$ parameter, with a
metallicity of $\gta 0.5Z_\odot$ up to solar.  The remaining two
objects are the two [NeIII] emitters, which are proven to be `normal'
star forming galaxies (rather than AGN) and confirmed to have the
intermediate, $Z \sim 0.3 Z_{\odot}$ value indicated, within the
errorbars, by the $R_{23}$ parameter.

The near-infrared data therefore demonstrate a high, i.e. about solar
metallicity for at least some objects in our $0.5<z<1$
high-$L_{H\beta}$ sample. The HST morphology of one of these
high-metallicity objects is that of a regular two-armed spiral (Schade
et al.\ 1995), and so it is possibly not surprising that this galaxy
has a high metal content. It may be that our high-$L_{H\beta}$
selection favours large well-formed galaxies relative to the general
blue CFRS population. However, there is no indication that this is the
case from the HST morphology of three additional objects for which the
HST data are available (Lilly \& Carollo 2001). More importantly, a
second object among the three with confirmed high metallicity and HST
morphology shows the irregularity and compactness typical of the blue
galaxy population at those redshifts.

Of course, until we obtain the infrared spectroscopy for the entire
sample, the $R_{23}$ degeneracy with $Z$ does not allow us to prove on
what branch ---i.e., the low- or the high-metallicity one--- each
individual high-$z$ galaxy lies, and the possibility remains that some
objects have indeed very low metallicities $Z\lta 0.1Z_\odot$, making
them strong candidates for star bursting dwarfs.  However, at this
stage, the confirmation that at least a large fraction of the
high-$L_{H\beta}$ $0.5 < z < 1$ star forming galaxies has high
metallicities within $\sim 40$-50\% of solar contrast the idea that
these systems are all low mass, i.e., low metallicity dwarfs
brightened by substantial luminosity evolution.  In contrast, the
metallicity data seem to suggest that at least a large fraction of
these small irregular galaxies are in fact the progenitors of today's
massive, metal-rich galaxies, but seen in an earlier phase of their
evolution when they were already significantly metal-rich but
morphologically more disturbed and smaller.

\end{document}